# High efficiency SHG of orbital angular momentum light in an external cavity

Zhi-Yuan Zhou[1,2,#1], Yan Li[1,2,#1], Dong-Sheng Ding[1,2], Yun-Kun Jiang[3], Wei Zhang[1,2], Shuai Shi[1,2], Bao-Sen Shi[1,2,*] and Guang-Can Guo[1,2]

[1]*Key Laboratory of Quantum Information, University of Science and Technology of China, Hefei, Anhui 230026, China*

[2]*Synergetic Innovation Center of Quantum Information & Quantum Physics, University of Science and Technology of China, Hefei, Anhui 230026, China*

[3]*College of Physics and Information Engineering, Fuzhou University, Fuzhou 350002, China*

[*]*Corresponding author: drshi@ustc.edu.cn*

Traditional methods for generating orbital angular momentum (OAM) light include holographic diffraction gratings, vortex phase plate and spatial light modulator. In this article, we report a new method for high efficient OAM light generation. By pumping an external cavity contains a quasi-phase matching nonlinear crystal with a fundamental OAM-carrying light and properly aligning the cavity, mode-matching between the pump light and the cavity's higher order Laguerre-Gaussian (LG) mode is achieved, conversion efficiency up to 10.3% have been obtained. We have demonstrated that the cavity can stably operate at its higher order LG mode just as Gaussian mode for the first time. The SHG light possesses a doubled OAM value with respect to the pump light. The parameters that affect the beam quality and conversion efficiency are discussed in detail. Our work opens a brand new field in laser optics, and makes the first step toward high efficiency OAM light processing.

Orbital angular momentum (OAM) light has grasped great attentions since first introduced by Allen et. al. [1], the singularity of the spatial structure of OAM light makes them suitable for optical trapping and manipulation [2, 3]; for there is no dimension limitation in the OAM degree of freedom of light, OAM-carrying light is preferred for high capacity optical communications [4, 5] and quantum key distribution [6]; high dimensional OAM entangled photon pairs are applied in quantum information processing for demonstrating basic principle of quantum mechanics [7-10]. Traditional methods for generating OAM light are holographic diffraction gratings, vortex phase plate and spatial light modulator. To generate OAM light with wavelengths at special regions such as in the UV and infrared regions, the above methods will be difficult. Frequency conversion of OAM light at wavelengths that are easier to obtain using nonlinear crystals offers the convenient

---

#1 These two authors have contributed equally to this article.

to achieve the above target. The frequency conversion of OAM light in birefringence phase matching (BPM) [11, 12] and quasi-phase matching (QPM) [13] crystals in the single pass configuration have been demonstrated. The above demonstrations only showed the possibility of frequency conversion of OAM light using nonlinear crystal, but the sufficient low conversion efficiency in the single pass configurations made them not suitable in practical usage. One common used method to enhance the conversion efficiency is to put the nonlinear crystal inside an optical cavity, the function of the cavity is to build up a circulating power that is many time relative to the pump power. For frequency conversion of pump light with Gaussian spatial shape, either BPM or QPM crystal is suitable. But for OAM light frequency conversion, QPM crystals have big advantages for its high effective nonlinear coefficient and no walk-off effect, the walk-off effect in BPM crystals distorts the frequency converted light spatial shape, and prevents one to obtain a high beam quality.

All previous researches on frequency conversion of light in an external cavity configuration are focused on Gaussian mode [14-17], no one else has study the frequency conversion of higher order mode in an external cavity. The OAM modes (or Lagurre-Gaussian modes) are also eigen modes of a confocal cavity, by properly aligning the cavity, it can on resonance with OAM mode similar as Gaussian mode. In this article, high conversion efficiency of light carries OAM of 2 in an external confocal cavity has been studied for the first time. We obtained 22.5mW of SHG light with pump power of 219 mW, a conversion efficiency of 10.3%, this efficiency is comparable with conversion efficiency with Gaussian pump light. When the cavity is locked, the leakage of the pump beam and the SHG beam all have a donut-like shapes, the OAM value for the SHG light is doubled with respect to the pump light, which was demonstrated using interferometer method showed in ref.[13]. The article is organized as follows: the experimental setup is described in the first part; the experiments results and the parameters that affect the conversion efficiency and beam quality are discussed in detail in the second part; then we come to the conclusion part, further improvements of the experiment, and the promising applications are discussed there.

The experimental setup is depicted in figure 1. The pump beam is from Ti:sapphire laser (Coherent, MBR110). The pump beam is modulated by an electro-optical modulator (EOM) with RF frequency of 11.9MHz; an isolator is placed behind the EOM to block back reflected light from the cavity; then the light is imprinted with OAM of value 2 using vortex phase plate (VPP, from RP photonics); the polarization of the OAM light is controlled using a half wave plate (HWP); the OAM light with proper polarization is mode-matched to the cavity using lenses; The confocal cavity is consisted of two concave mirrors (CM1 and CM2); the input coupler CM1 of curvature 51.85mm, has a transmittance of 4% for the pump beam at 795nm; the output coupler CM2 of curvature 60mm is high reflected coated for 795nm (R >99.9%) and high transmittance coated at 397.5nm (T>95%). The total length of the cavity is 80mm. The periodically poled KTP (PPKTP) crystal (Raicol Crystals) has a dimension of 1mm×2mm×10mm, both end faces are anti-reflected coated at 795nm and 397.5nm. The temperature of the crystal is controlled with a homemade temperature controller with stability of ±2mK; A piezoelectric transducer (PZT) is attached to CM2 for scanning and locking the cavity. The leakage of the pump beam and the generated SHG beam are separated by dichromatic mirror (DM), the transmission spectral of the cavity is detected using a fast photodiode (D). The signal from the detector is mixed with modulation signal of the EOM and filtered using a low pass filter to generate error signal for locking cavity. The error signal is processed using a homemade PID circuit and amplified using a high voltage amplifier to control the PZT.

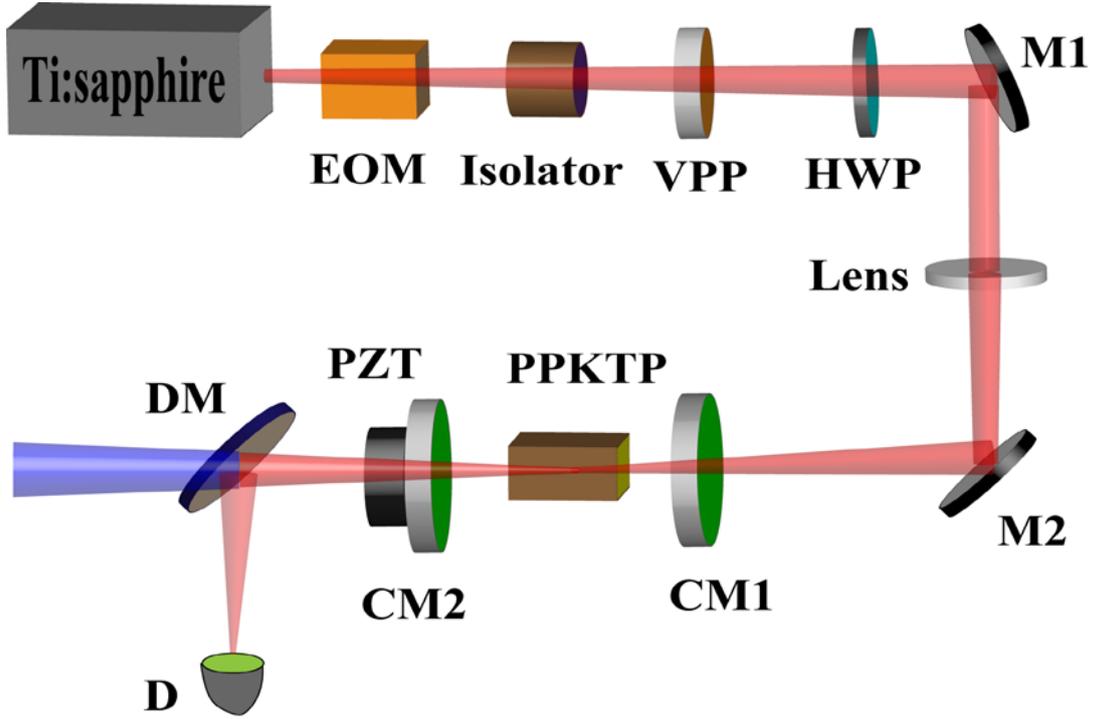

Figure 1. Experimental setup for our experiments.

The experimental results are showed in figure 2. When the cavity is aligned properly on resonance with the pump OAM beam, the transmission spectrum detected by photodiode and recorded using oscilloscope is showed in figure 2d, the highest peak corresponding to the LG eigen mode of the cavity, the other small peaks are unrelated mode due to imperfect mode matching between the pump beam and the cavity. When the cavity is locked to the highest peak using Pound-Drever-Hall method (PDH) [18], the leakage of the pump beam from the output coupler has a donut shape shows in figure 2a, the generated SHG beam has a donut spatial shape shows in figure 2b. The asymmetry and distortion in the shape is arising from defects in the cavity mirror and aberrations of beam propagation inside the cavity. We have demonstrated in our previous work [13] that the OAM is conversation in the SHG process, so the SHG beam possesses OAM value twice as that of the pump OAM value. We send the SHG beam into an interferometer the same as in Ref. [13], the interference pattern has a flower-like shape with 8 petals, so the OAM value of the SHG beam is 4, which is consistent with previous study. We also measure the SHG power as function of the pump power. The result is plotted in figure 2e, it shows that the SHG power is increasing nearly squared with respect to the increasing of the pump power, which is similar to SHG using Gaussian mode. When the pump power is 219mW, we obtained a SHG power of 22.5mW, corresponding to single side efficiency of 10.3%, if the output at the other mirror is considered, the efficiency will reach 20.6%, this efficiency is comparable with SHG pumping using Gaussian mode. For SHG using Gaussian mode, there exist an optimal coupling coefficient of the input coupler [16], this will be also true for SHG pump using LG mode. The locking behavior of our cavity on resonance with LG mode is the same as Gaussian mode, our cavity can be operated on locking stably for hours.

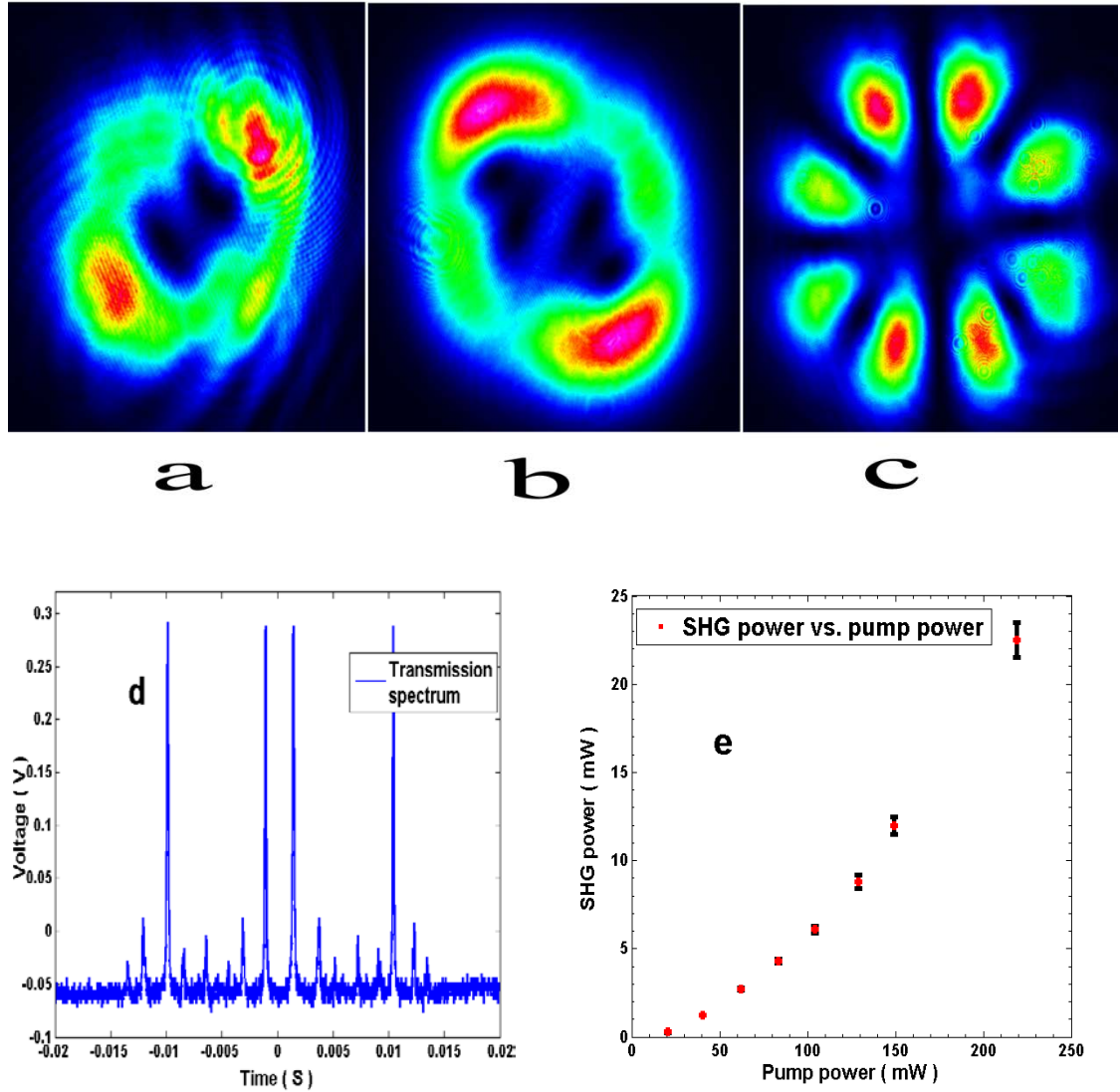

Figure 2. Experimental results. Images a and b are the shape of the pump beam leakage from the cavity output coupler and the shape of the SHG beam when the cavity is on locking respectively; image c is interference pattern of the SHG beam by sending it to an interferometer shows in ref. [13]; d is the transmission spectrum of the cavity; e shows the SHG power as function of the pump power.

In conclusion, we have realized high efficiency SHG in an external cavity pumping using OAM light for the first time. Both the leakage of the pump and the SHG beams have a donut-like shapes, the OAM value of the SHG beam is verified to be doubled with respect to the pump. In this very first study, the mode-matching is not perfect and the beam quality is not that much so good, these problems can be solved by using high quality mirrors and purifying the spatial mode from the laser using single mode fiber. The conversion efficiency can also be improved by choosing a proper input coupling coefficient and designing a cavity with smaller beam waist. In our future study, we will convert OAM light base on sum frequency generation in an external cavity, in this configuration, light with arbitrary OAM value can be converted with high efficient. The present work opens a new field in laser optics, and makes the first step towards high efficiency OAM light processing, which will find applications in OAM light based high capacity optical communications and quantum information processing.